\documentclass[11pt,a4paper]{article}

\usepackage{epsfig,amsmath,amssymb,cite}

\begin{document}

\title{Spherical Skyrmion black holes as gravitational lenses}
\author{Fabrizio Canfora$^{1,}$\thanks{e-mail: canfora@cecs.cl}, 
Ernesto F. Eiroa$^{2,3,}$\thanks{e-mail: eiroa@iafe.uba.ar}, 
Carlos M. Sendra$^{2,3,}$\thanks{e-mail: cmsendra@iafe.uba.ar} \\
{\small $^1$ Centro de Estudios Cient\'{\i}ficos (CECS), Casilla 1469, Valdivia, Chile}\\
{\small $^2$ Instituto de Astronom\'{\i}a y F\'{\i}sica del Espacio (IAFE, CONICET-UBA),} \\
{\small Casilla de Correo 67, Sucursal 28, 1428, Buenos Aires, Argentina}\\
{\small $^3$ Departamento de F\'{\i}sica, Facultad de Ciencias Exactas y Naturales, Universidad de Buenos Aires, } \\
{\small Ciudad Universitaria Pabell\'on I, 1428, Buenos Aires, Argentina} }
\maketitle
\date{}

\begin{abstract}
In this article, we extend the strong deflection limit to calculate the deflection angle for a class of geometries which are asymptotically locally flat. In particular, we study the deflection of light in the surroundings of spherical black holes in Einstein-Skyrme theory. We find the deflection angle in this limit, from which we obtain the positions and the magnifications of the relativistic images. We compare our results with those corresponding to the Schwarzschild and the global monopole (Barriola-Vilenkin) spacetimes.
\end{abstract}

\section{Introduction}\label{intro}

The presence of supermassive black holes at the center of most galaxies, in particular the Milky Way \cite{gillessen17} and the closest one M87 \cite{broderick15}, has led to a growing interest in the optical effects in their neighborhood. It is believed that the observation of some of these effects, including direct imaging, will be possible in the near future \cite{observ}. Regarding gravitational lensing by compact objects possessing a photon sphere --like black holes--, besides the primary and secondary images, there exist two infinite sets of the denominated relativistic images \cite{virbha1}, produced by light rays passing close to the photon sphere, then having large deflection angles. In this case, the deflection angle admits a logarithmic approximation dubbed the \textit{strong deflection limit}, which allows for obtaining analytically the positions, the magnifications, and the time delays of the relativistic images. This approximate procedure was firstly introduced for the Schwarzschild black hole \cite{darwin-otros}, extended to the Reissner--Nordstr\"{o}m spacetime \cite{eiroto}, and then generalized to any spherically symmetric and asymptotically flat geometries \cite{bozza02}. This method was recently simplified and improved \cite{tsukamoto}. The continued advances in gravitational lensing observations has lead to a growing interest in the analysis of lensing effects as a possible test of gravitational theories in the strong regime. Many analysis of strong deflection lenses have been considered in recent years \cite{nakedsing1,nakedsing2,retro,lenseq,virbha2,alternative,bwlens,scalarlens,
cm,tsukamoto-wh} both within general relativity as well as in alternative theories of gravity. The lensing effects of rotating black holes have also been considered in the literature \cite{rotbh}. In this case, the deformation of the shadow is another related topic of great interest \cite{shadows1,shadows2}.

Within this context, it is relevant the study of strong deflection gravitational lensing when General Relativity is coupled with the action describing the strong interactions of baryons and mesons. Such an action corresponds to Skyrme's theory \cite{skyrme} (detailed reviews are \cite{revs}) which describes the low energy limit of QCD \cite{witten0,bala0}. The dynamical variable of the Skyrme action is a scalar field $U$ taking value in $SU(N)$ (here we will consider the $SU(2)$ case). The agreement of the theoretical predictions of Skyrme theory with experiments is very good (a partial list of relevant references is \cite{spin,revs,bala0,bala12,ANW,mgm,skyrmonoratio} and references therein). Due to these reasons, the Einstein-Skyrme system has attracted a lot of attention. In a series of seminal papers Droz, Heusler, and Straumann \cite{3} (following the findings of Luckock and Moss \cite{4}) constructed black hole solutions with a non-trivial Skyrme hair with a spherically symmetric ansatz. The issue of linear stability has also been analyzed in \cite{5}. On the other hand, until very recently, there were basically no analytic solution in the Einstein-Skyrme system. Here we want to remark that the search for analytic configurations in the Skyrme and Einstein-Skyrme theories is not just of academic interest\footnote{For instance, in the Skyrme model in flat spaces, it was well known that if one includes a too large isospin chemical potential then the Skyrmion becomes unstable. However, only very recently in \cite{CG23} it was derived an analytic formula for this critical chemical potential.}. Using some recent results on the generalization of the hedgehog ansatz \cite{56,58,yang12,59,60} an analytic spherically symmetric black hole in the $SU(2)$ Einstein-Skyrme theory has been constructed \cite{46}. In this black hole, the effects of the Skyrme are manifest and so it offers the intriguing possibility to analyze a gravitational lens which includes the effects of strong interactions. Such a possibility will be explored in the present paper.

The article is organized as follows. In Sec. \ref{skyrme}, we introduce the $U(2)$ Einstein-Skyrme system and we present the spherically symmetric black hole solution. In Sec. \ref{sdl}, we extend the strong deflection limit for the deflection angle to a class of spherically symmetric spacetimes which are not necessarily asymptotically Minkowski. In Sec. \ref{images}, we obtain the angular positions and the magnifications of the relativistic images. In Sec. \ref{skirmlens}, we apply the method to the Skyrmion black hole. Finally, in Sec. \ref{discu}, we analyze the results obtained. We adopt Planck units, so that $G=c=\hslash =1$.

\section{The $SU(2)$ Einstein - Skyrme system}\label{skyrme}

This section is a very standard and short review of the Einstein-Skyrme action. The $SU(2)$ Skyrme field is a $SU(2)$-valued scalar field so that the Einstein-Skyrme action is described by 
\begin{equation}
S=S_{\mathrm{G}}+S_{\mathrm{Skyrme}},
\end{equation}
where the gravitational action $S_{\mathrm{G}}$ and the Skyrme action $S_{\mathrm{Skyrme}}$ are given by 
\begin{align}
S_{\mathrm{G}}=& \frac{1}{16\pi }\int d^{4}x\sqrt{-g}(\mathcal{R}-2\Lambda ),
\label{einskyrm} \\
S_{\mathrm{Skyrme}}=& \int d^{4}x\sqrt{-g}\mathrm{Tr}\left( \frac{F_{\pi
}^{2}}{16}R^{\mu }R_{\mu }+\frac{1}{32e^{2}}F_{\mu \nu }F^{\mu \nu }\right) .  \label{sky}
\end{align}
Here $R_{\mu }$ and $F_{\mu \nu }$ are defined by 
\begin{align}
R_{\mu }=& U^{-1}\nabla _{\mu }U\ ,  \label{skyrme2} \\
F_{\mu \nu }=& \left[ R_{\mu },R_{\nu }\right] ,   \label{skyrmenotation}
\end{align}
while the positive parameters $F_{\pi }$ and $e$ are fixed by comparison with experimental data. The Skyrme fields satisfy the dominant energy condition \cite{Gibb2003}.

For convenience, we define $K=F_{\pi }^{2}/4$ and $\lambda = 4/(e^{2}F_{\pi}^{2})$, where (see the discussions in \cite{ANW,58}) 
\begin{equation}
F_{\pi }=141\ MeV,\; 5\leq e\leq 7 .  \label{window}
\end{equation}
Indeed, it is well known that while the parameter $F_{\pi }$ can be determined precisely by comparison with nuclear spectra, there is some uncertainty on the parameter $e$. Thus, all the values of the parameter $e$ in the above window can be considered as reasonable.

The Skyrme action can be written as 
\begin{equation}
S_{\mathrm{Skyrme}}=\frac{K}{2}\int d^{4}x\sqrt{-g}\mathrm{Tr}\left( \frac{1}{2}R^{\mu }R_{\mu }+\frac{\lambda }{16}F_{\mu \nu }F^{\mu \nu }\right) \ .
\label{skyrme-action}
\end{equation}
The resulting Einstein equations are 
\begin{equation}
G_{\mu \nu }+\Lambda g_{\mu \nu }=8\pi T_{\mu \nu },  \label{einskyrmequ}
\end{equation}
where $G_{\mu \nu }$ is the Einstein tensor and 
\begin{equation}
T_{\mu \nu }=-\frac{K}{2}\mathrm{Tr}\biggl[\biggl(R_{\mu }R_{\nu }-\frac{1}{2}g_{\mu \nu }R^{\alpha }R_{\alpha }\biggl)+\frac{\lambda }{4}\biggl(g^{\alpha \beta }F_{\mu \alpha }F_{\nu \beta }-\frac{1}{4}g_{\mu \nu
}F_{\alpha \beta }F^{\alpha \beta }\biggl)\biggl] .  \label{timunu1}
\end{equation}
The Skyrme equations are written as 
\begin{equation}
\nabla ^{\mu }R_{\mu }+\frac{\lambda }{4}\nabla ^{\mu }[R^{\nu },F_{\mu \nu
}]=0\ .  \label{nonlinearsigma1}
\end{equation}
Here $R_{\mu }$ is expressed as 
\begin{equation}
R_{\mu }=R_{\mu }^{i}\tau _{i} ,
\end{equation}
in the basis of the SU(2) generators 
\begin{equation*}
\tau ^{k}=i\sigma ^{k} ,
\end{equation*}
(where $\sigma ^{k}$ are the Pauli matrices, the Latin index $i=1,2,3$ corresponds to the group index, which is raised and lowered with the flat metric $\delta_{ij}$), which satisfy 
\begin{equation}
\tau ^{i}\tau ^{j}=-\delta ^{ij}\mathbf{1}-\varepsilon ^{ijk}\tau ^{k} ,
\end{equation}
where $\mathbf{1}$ is the identity $2\times 2$ matrix and $\varepsilon _{ijk}$ and $\varepsilon ^{ijk}$ are the totally antisymmetric Levi-Civita symbols with $\varepsilon _{123}=\varepsilon ^{123}=1$.

Hereafter we will use the following standard parametrization of the SU(2)-valued scalar $U(x^{\mu })$:
\begin{equation}
U(x^{\mu })=Y^{0}\mathbf{1}+Y^{i}\tau _{i}\ ,\quad U^{-1}(x^{\mu })= Y^{0} 
\mathbf{1}-Y^{i}\tau _{i} ,  \label{standard1}
\end{equation}
where $Y^{0}=Y^{0}(x^{\mu })$ and $Y^{i}=Y^{i}(x^{\mu })$ satisfy 
\begin{equation}
\left( Y^{0}\right) ^{2}+Y^{i}Y_{i}=1 .  \label{standard3}
\end{equation}
From the definition~(\ref{skyrme2}), $R_{\mu }^{k}$ is written as 
\begin{equation}
R_{\mu }^{k}=\varepsilon ^{ijk}Y_{i}\nabla _{\mu }Y_{j}+Y^{0}\nabla _{\mu
}Y^{k}-Y^{k}\nabla _{\mu }Y^{0} .  \label{standard4}
\end{equation}
Using the quadratic combination 
\begin{equation}
\mathcal{S}_{\mu \nu }:=\delta _{ij}R_{\mu }^{i}R_{\nu }^{j}=G_{ij}(Y)\nabla
_{\mu }Y^{i}\nabla _{\nu }Y^{j} ,  \label{cuadra1}
\end{equation}
where 
\begin{equation}
G_{ij}:=\delta _{ij}+\frac{Y_{i}Y_{j}}{1-Y^{k}Y_{k}}\ ,
\end{equation}
we obtain 
\begin{equation}
\mathrm{Tr}(R_{\mu }R_{\nu })= -2\mathcal{S}_{\mu \nu }, \qquad \mathrm{Tr}
(F_{\mu \alpha }F_{\nu }^{~\alpha })= 8\mathcal{S}_{\mu \alpha } 
\mathcal{S}_{\nu }^{~\alpha }-8\mathcal{S}_{\mu \nu }\mathcal{S}.
\end{equation}
Using these results, we can write the Skyrme action (\ref{skyrme-action}) only with $Y^{i}$ as 
\begin{align}
S_{\mathrm{Skyrme}}=& -K\int d^{4}x\sqrt{-g}\left\{ \frac{1}{2}G_{ij}(\nabla
_{\mu }Y^{i})(\nabla ^{\mu }Y^{j}) +\frac{\lambda }{4}\left[ \left(
G_{ij}\left( \nabla _{\mu }Y^{i})(\nabla ^{\mu }Y^{j}\right )\right) ^{2}
\right. \right.  \notag \\
& \left. \left. -G_{ij}(\nabla _{\mu }Y^{i})(\nabla _{\nu
}Y^{j})G_{kl}(\nabla ^{\mu }Y^{k})(\nabla ^{\nu }Y^{l})\right] \right\},
\end{align}
while the energy-momentum tensor (\ref{timunu1}) is expressed as 
\begin{equation}
T_{\mu \nu }= K\left\{ \mathcal{S}_{\mu \nu }-\frac{1}{2}g_{\mu \nu } 
\mathcal{S}+\lambda \left[ \mathcal{S}\mathcal{S}_{\mu \nu }-\mathcal{S}
_{\mu \alpha }\mathcal{S}_{\nu }^{~\alpha } -\frac{1}{4}g_{\mu \nu }\left( 
\mathcal{S}^{2}-\mathcal{S}_{\alpha \beta } \mathcal{S}^{\alpha \beta
}\right) \right] \right\} .   \label{tmunu2}
\end{equation}

The field equations admit a spherically symmetric solution, which represents a spherical black hole in Einstein-Skyrme theory \cite{46} 
\begin{equation}
ds^{2}=-f(r)dt^{2}+f(r)^{-1}dr^{2}+r^{2}(d\vartheta ^{2}+\sin ^{2}\vartheta d\varphi ^{2}) ,  \label{sm1}
\end{equation}
where the metric functions are given by 
\begin{equation}
f(r)=1-8\pi K-\frac{2M}{r}+\frac{4\pi K\lambda }{r^{2}}-\frac{1}{3}\Lambda
r^{2} .  \label{sm2}
\end{equation}
The Skyrme source for the above metric corresponds to take in Eqs. (\ref{standard1}) and (\ref{standard3}) 
\begin{equation}
Y_{0}=0\ ,\quad Y_{1}=\sin \vartheta \cos \varphi \ ,\quad Y_{2}=\sin \vartheta \sin \varphi \ ,\quad Y_{3}=\cos \vartheta  .  \label{MF}
\end{equation}
Indeed, one can verify easily that the metric in Eqs. (\ref{sm1}) and (\ref{sm2}) and the Skyrme field in Eq. (\ref{MF}) solve the coupled field equations (\ref{einskyrmequ}) and (\ref{nonlinearsigma1}) with the energy-momentum tensor in Eq. (\ref{timunu1}).

It is interesting to note that, when $\Lambda =0$, the above spherical black hole in Einstein-Skyrme theory can be interpreted as the black hole of Barriola-Vilenkin type \cite{BVM} (since $f(r) \rightarrow 1-8\pi K<1$ when $r\rightarrow \infty $) but in which the Skyrme coupling $\lambda $ gives an explicit contribution to $f(r)$ of order $1/r^{2}$. The role of this term will be apparent in the following analysis.

\section{Deflection angle in the strong deflection limit}\label{sdl}

We start by adopting the spherically symmetric geometry 
\begin{equation}
ds^{2}=-A(r)dt^{2}+B(r)dr^{2}+C(r)(d\vartheta ^{2}+\sin ^{2}\vartheta d\varphi ^{2}),  \label{m1}
\end{equation}
where the metric functions satisfy 
\begin{equation}
\lim_{r\rightarrow \infty }A(r)=\lim_{r\rightarrow \infty }B(r)^{-1}=\mu \
,\quad \lim_{r\rightarrow \infty }C(r)=r^{2},  \label{m3}
\end{equation}
with $\mu $ a positive constant. When $\mu \neq 1$ in the above equations, the corresponding spherical geometry is dubbed as \textit{asymptotically locally flat}. The most famous example of an asymptotically locally flat is the Barriola-Vilenkin metric \cite{BVM}\ which describes the space-time of a global monopole. This means that we are interested in extending the strong deflection limit to the asymptotically locally flat scenario, i.e. in which the functions $A(r)$ and $B(r)^{-1}$ approach a positive constant when $r\rightarrow \infty $, but this constant is not necessarily the number $1$ as in the usual case of Minkowski asymptotics\footnote{After a suitable coordinate change, it is easy to see that there is a solid angular deficit if $\mu <1$ or a solid angular surplus if $\mu >1$. This situation is common when topological defects (such as cosmic strings or in the Skyrme model) are present.}. The radius of the event horizon $r_{h}$ is given by the largest root of the function $A(r)$. We assume in what follows that all the metric functions are positive and finite for $r>r_{h}$. The photon sphere corresponds to an unstable circular orbit for massless particles. We define 
\begin{equation}
D(r)=\frac{C^{\prime }(r)}{C(r)}-\frac{A^{\prime }(r)}{A(r)},  \label{d}
\end{equation}
where the prime symbol denotes differentiation with respect to the coordinate $r$. We assume that the equation $D(r)=0$ has at least a positive solution, being the radius of the photon sphere $r_{m}$ the largest one of them.

Let us consider a photon coming from infinity, reaching the closest approach distance $r_0>r_m$, and returning to infinity. Due to the symmetries of geometry (\ref{m1}), the null geodesics have two conserved quantities $E$ (energy) and $L$ (angular momentum), and the movement is confined to a plane, which can be taken with constant $\vartheta = \pi /2$, without losing generality. By parameterizing the trajectory with an affine parameter, it is straightforward to verify that 
\begin{equation}
-A(r)\dot{t}^2+B(r)\dot{r}^2+C(r)\dot{\varphi}^2=0,  \label{null}
\end{equation}
with the dot symbol representing the derivative with respect to the affine parameter. As usual, by combining this equation with 
\begin{equation}
E=A(r)\dot{t}  \label{ener}
\end{equation}
and 
\begin{equation}
L=C(r)\dot{\varphi},  \label{L}
\end{equation}
we can obtain the radial equation 
\begin{equation}
\dot{r}^2=V(r),  \label{req}
\end{equation}
in terms of the effective potential 
\begin{equation}
V(r)=\frac{L^2 R(r)}{B(r) C(r)},  \label{pot}
\end{equation}
where 
\begin{equation}
R(r)=\frac{C(r)}{u^2A(r)}-1,
\end{equation}
with $u=L/E$ the impact parameter. The photon is allowed to move in the region with $V(r)\ge 0$. By using the asymptotic condition (\ref{m3}), we can easily see that $ V(r) \rightarrow E^2 > 0$ when $r\rightarrow \infty$, so the photon can exist at an infinite radius. The assumption of the existence of a closest approach radius $r_0$ implies that $R(r)=0$ should have at least one positive solution. At the point with $r=r_0$, we have that $\dot{r}=0$, so by using Eqs. (\ref{null}), (\ref{ener}), and (\ref{L}), we obtain that 
\begin{equation}
u=\sqrt{\frac{C_0}{A_0}},  \label{impactu1}
\end{equation}
and then 
\begin{equation}
R(r)=\frac{A_0 C(r)}{A(r) C_0}-1,
\end{equation}
where, here and from now on, the subscript $0$ stands for evaluation at $r=r_0$ in the metric functions. From Eqs. (\ref{L}), (\ref{req}), and Eq. (\ref{pot}), we find that the trajectory is determined by 
\begin{equation}
\left( \frac{dr}{d\varphi} \right) ^2 =\frac{R(r) C(r)}{B(r)},
\end{equation}
so, by integrating this equation, the deflection angle for a photon coming from infinity can be written in the same form as in Refs. \cite{weinberg,nakedsing1} 
\begin{equation}
\alpha(r_0)=I(r_0)-\pi,  \label{alfa1}
\end{equation}
with 
\begin{equation}
I(r_0)=2\int^{\infty}_{r_0}\frac{\sqrt{B(r)}}{\sqrt{R(r)C(r)}}dr.  \label{i0}
\end{equation}
The deflection angle is a monotonic decreasing function of the closest approach distance $r_0$. There is a logarithmic divergence in $\alpha$ as $r_0 $ approaches to the photon sphere radius $r_m$. For smaller values than a certain $r_0$, the deflection angle becomes greater than $2\pi$, which means that the photons perform more than one turn around the black hole before they emerge from it. This gives place to two infinite sets of relativistic images, one at each side of the black hole, which can be studied by performing the strong deflection limit. By replacing the metric functions in Eq. (\ref{alfa1}), the exact deflection angle can be obtained, in most cases numerically, but in few ones, analytically. In order to find an analytic expression of the deflection angle in the strong deflection limit, the integral (\ref{i0}) can be suitably rewritten by following the procedure introduced in Refs. \cite{bozza02, tsukamoto}. By making the change of variables 
\begin{equation}
z\equiv 1-\frac{r_0}{r},  \label{zvariable}
\end{equation}
the integral (\ref{i0}) takes the form 
\begin{equation}
I(r_0)=\int^{1}_{0}f(z,r_0)dz,  \label{i02}
\end{equation}
where 
\begin{equation}
f(z,r_0)=\frac{2r_0}{\sqrt{G(z,r_0)}},  \label{fdef}
\end{equation}
being 
\begin{equation}
G(z,r_0)\equiv R(r)\frac{C(r)}{B(r)}(1-z)^4.  \label{gdef}
\end{equation}
It is convenient to split the integral (\ref{i0}) in two parts 
\begin{equation}
I(r_0)=I_D(r_0)+I_R(r_0),  \label{i03}
\end{equation}
where $I_D(r_0)$ contains the divergence at $r_0=r_m$ and $I_R(r_0)$ is regular everywhere. The divergent part can be written as 
\begin{equation}
I_D(r_0)\equiv\int^{1}_{0}f_D(z,r_0)dz,  \label{id}
\end{equation}
with $f_D(z,r_0)$ given by 
\begin{equation}
f_D(z,r_0)\equiv\frac{2r_0}{\sqrt{c_1(r_0)z+c_2(r_0)z^2}},  \label{fd}
\end{equation}
where 
\begin{equation}
c_1(r_0)=\frac{C_0 D_0 r_0}{B_0}  \label{c1}
\end{equation}
and 
\begin{equation}
c_2(r_0)=\frac{C_0 r_0}{B_0}\left\{D_0\left[\left(D_0-\frac{B^{\prime }_0}{B_0}\right)r_0-3\right]+\frac{r_0}{2}\left(\frac{C^{\prime \prime }_0}{C_0}- 
\frac{A^{\prime \prime }_0}{A_0}\right)\right\}.  \label{c2}
\end{equation}
For photons passing close to the photon sphere, the strong deflection limit is performed by taking $r_0\rightarrow r_m$. In this limit, Eq. (\ref{d}) results $D(r_m)=0$, and expressions (\ref{c1}) and (\ref{c2}) reduces to zero 
\begin{equation}
c_1(r_m)=0
\end{equation}
and 
\begin{equation}
c_2(r_m)=\frac{C_m r^{2}_{m}}{2B_m}\left(\frac{C^{\prime \prime }_m}{C_m} -\frac{A^{\prime \prime }_m}{A_m}\right),  \label{c2m}
\end{equation}
where the subscript $m$ denotes evaluation in $r=r_m$ in the corresponding functions. The integral $I_R$ is defined by 
\begin{equation}
I_R(r_0)\equiv\int^{1}_{0}f_R(z,r_0)dz,  \label{ir}
\end{equation}
with 
\begin{equation}
f_R(r_0)\equiv f(z,r_0)-f_D(z,r_0),  \label{fr}
\end{equation}
which is regular since it has the divergence subtracted. In terms of the impact parameter $u$, the deflection angle in the strong deflection limit is given by 
\begin{equation}
\alpha(u)=-a_1\ln\left(\frac{u}{u_m}-1\right)+a_2+O((u-u_m)
\ln(u-u_m)),  \label{alpha}
\end{equation}
where $a_1$ and $a_2$ are the so called strong deflection limit coefficients, which depend only on the metric functions, as follow: 
\begin{equation}
a_1=\sqrt{\frac{2B_m A_m}{C^{\prime \prime }_m A_m-C_mA^{\prime \prime
}_m}}  \label{abarra1}
\end{equation}
and 
\begin{equation}
a_2=a_1\ln\left[r^{2}_{m}\left(\frac{C^{\prime \prime }_m}{C_m}- 
\frac{A^{\prime \prime }_m}{A_m}\right)\right]+I_R(r_m)-\pi.  \label{bbarra1}
\end{equation}
The critical impact parameter $u_m$ corresponds to photons with $r_0 \rightarrow r_m$. The approximate expression (\ref{alpha}) for the deflection angle is the starting point for the analytical calculation of the positions and the magnifications of the relativistic images.

\section{Relativistic images}\label{images}

We consider the case where a source of light is behind a black hole lens, with the optical axis defined as the line joining the lens and the observer. We assume that the observer-lens $D_{ol}$ and the lens-source $D_{ls}$ angular diameter (coordinate) distances are much greater than the horizon radius $r_{h}$, and that the observer-source distance satisfies $D_{os} = D_{ol}+D_{ls}$. The deflection of the photons takes place in the small region close to the black hole; far away from it our geometry is locally flat so, as in the usual case of Minkowski asymptotics, the trajectories of photons can be approximated by straight lines. Then, we adopt a lens equation that can be written in the same form as the one presented in Ref. \cite{lenseq}:
\begin{equation}
\tan \beta =\frac{D_{ol}\sin \theta -D_{ls}\sin (\alpha -\theta )}
{D_{os} \cos (\alpha -\theta )},  \label{pm1}
\end{equation}
where $\beta $ is the angular position of the source and $\theta$ the angular position of an image detected by the observer, both taken from the optical axis. When the objects are highly aligned, the lensing effects are more relevant. In this situation, the angles $\beta $ and $\theta $ are small, $\alpha $ is close to a multiple of $2\pi $, and two infinite sets of point relativistic images are obtained if $\beta \neq 0$. The deflection angle can be written as $\alpha =\pm 2n\pi \pm \Delta \alpha _{n}$, with $n\in \mathbb{N}$ and $0<\Delta \alpha _{n}\ll 1$, where the $+/-$ sign corresponds to first/second set of relativistic images. By replacing this deflection angle in Eq. (\ref{pm1}), we have 
\begin{equation}
\beta =\theta \mp \frac{D_{ls}}{D_{os}}\Delta \alpha _{n},  \label{pm2}
\end{equation}
which is the same equation obtained in Refs. \cite{bozza02,lenseq} for Minkowski asymptotics. Here, the $-/+$ sign stands for the first/second set of relativistic images. The impact parameter results $u=D_{ol}\sin \theta \approx D_{ol}\theta $ from geometrical considerations, and the deflection angle equation (\ref{alpha}) takes the form 
\begin{equation}
\alpha (\theta ) = -a_1\ln \left( \frac{D_{ol}\theta }{u_{m}}
-1\right) +a_2.  \label{pm4}
\end{equation}
By inverting Eq. (\ref{pm4}) and performing a first order Taylor expansion around $\alpha =2n\pi $, we obtain the angular position of the $n$-th image, which for the first set of relativistic images results 
\begin{equation}
\theta _{n}=\theta _{n}^{0}-\zeta _{n}\Delta \alpha _{n},  \label{pm6}
\end{equation}
where 
\begin{equation}
\theta _{n}^{0}=\frac{u_{m}}{D_{ol}}\left[ 1+e^{(a_2-2n\pi )/a_1}
\right]  \label{pm7}
\end{equation}
and 
\begin{equation}
\zeta _{n}=\frac{u_{m}}{a_1 D_{ol}}e^{(a_2-2n\pi )/a_1}.
\label{pm8}
\end{equation}
Replacing $\theta _{n}$ in Eq. (\ref{pm2}), we have $\Delta \alpha _{n}=(\theta _{n}-\beta )D_{os}/D_{ls}$, and putting this expression in Eq. (\ref{pm6}), results in 
\begin{equation}
\theta _{n}=\theta _{n}^{0}-\frac{\zeta _{n}D_{os}}{D_{ls}}(\theta
_{n}-\beta ).  \label{pm10}
\end{equation}
Considering $0<\zeta _{n}D_{os}/D_{ls}<1$ and keeping only the first-order term in $\zeta _{n}D_{os}/D_{ls}$, the angular positions of the images finally take the form 
\begin{equation}
\theta _{n}=\theta _{n}^{0}+\frac{\zeta _{n}D_{os}}{D_{ls}}(\beta -\theta_{n}^{0}).  \label{pm14}
\end{equation}
For the other set of the relativistic images, we obtain analogously 
\begin{equation}
\theta _{n}=-\theta _{n}^{0}+\frac{\zeta _{n}D_{os}}{D_{ls}}(\beta +\theta
_{n}^{0}).  \label{pm15}
\end{equation}

The magnification of the $n$-th relativistic image is defined by the quotient of the solid angles subtended by the image and the source 
\begin{equation}
\mu_n=\left|\frac{\sin\beta}{\sin\theta_n}\frac{d\beta}{d\theta_n}
\right|^{-1},  \label{mag1}
\end{equation}
which, considering small angles and using Eq. (\ref{pm14}), we obtain 
\begin{equation}
\mu _{n}=\frac{1}{\beta}\left[ \theta ^{0}_{n}+ \frac {\zeta _{n}D_{os}}{
D_{ls}}(\beta - \theta ^{0}_{n})\right] \frac {\zeta _{n}D_{os}}{D_{ls}},
\label{pm18}
\end{equation}
and performing a first order Taylor expansion in $\zeta_n D_{os}/D_{ls}$, we finally have for both set of relativistic images 
\begin{equation}
\mu _{n}=\frac{1}{\beta}\frac{\theta ^{0}_{n}\zeta _{n}D_{os}}{D_{ls}}.
\label{pm19}
\end{equation}
The first relativistic image is the brightest one since the magnifications decreases exponentially with $n$. All images are very faint because their magnifications are proportional to $(u_{m}/D_{ol})^2$.

From the positions and the magnifications of the relativistic images, the following observables can be defined \cite{bozza02}: 
\begin{equation}
\theta_{\infty}=\frac{u_{m}}{D_{ol}},  \label{ob1}
\end{equation}
\begin{equation}
s=\theta_1-\theta_{\infty},  \label{ob2}
\end{equation}
and 
\begin{equation}
r=\frac{\mu_1}{\sum^{\infty}_{n=2}\mu_n},  \label{ob3}
\end{equation}
where $s$ corresponds to the angular separation between the position of the first relativistic image and the limiting value of the others $\theta_{\infty}$, and $r$ is the quotient between the flux of the first image and the flux coming from all the other images. For high alignment, these observables take the simple form \cite{bozza02}: 
\begin{equation}
s=\theta_{\infty}e^{(a_2-2\pi)/a_1}  \label{ob4}
\end{equation}
and 
\begin{equation}
r=e^{2\pi/a_1},  \label{ob5}
\end{equation}
which depend on the geometry of the black hole since they are functions of the strong deflection limit parameters.

\section{Application to the Skyrmion black hole}\label{skirmlens}

In this work, we are interested in the asymptotically locally flat case (i.e. without the cosmological constant) of the Skyrmion spacetime, so we take $\Lambda =0$ in Eq. (\ref{sm2}). Then, the metric functions take the form 
\begin{equation}
A(r)=B(r)^{-1}=1-8\pi K-\frac{2M}{r}+\frac{4\pi K\lambda }{r^{2}}\ ,\quad
C(r)=r^{2}\ .  \label{m2}
\end{equation}
These metric functions satisfy the condition (\ref{m3}) by identifying $\mu = 1-8\pi K$ and assuming that $8\pi K<1$. The radius of the event horizon for the spherical black hole of Einstein-Skyrme theory defined above, corresponding to the largest solution of $A(r)=0$, results 
\begin{equation}
r_{h}=\frac{M+\sqrt{M^{2}-4\pi K\lambda (1- 8\pi K) }}{1-8\pi K},  \label{xh}
\end{equation}
while, by using Eq. (\ref{d}), the photon sphere radius has the form 
\begin{equation}
r_{m}=\frac{3M+\sqrt{9M^{2}-32\pi K\lambda (1 - 8\pi K) }}{2(1-8\pi K)}.
\label{rm}
\end{equation}
We require that the photon sphere is always present, so the condition $32\pi K\lambda -256\pi ^{2}K^{2}\lambda \leq 9M^{2}$ should be satisfied. By performing the calculations in Eq. (\ref{ir}), we obtain the regular part of the integral 
\begin{equation}
I_{R}(r_{m})=\frac{-2\,r_{m}}{\sqrt{3Mr_{m}-16\pi K\lambda }}\ln \left[ 
\frac{2Mr_{m}-8\pi K\lambda +\sqrt{\left( 3Mr_{m}-16\pi K\lambda \right)
\left( Mr_{m}-4\pi K\lambda \right) }}{6Mr_{m} - 32\pi K\lambda }\right] .
\end{equation}
By replacing the metric functions in Eq. (\ref{impactu1}), the impact parameter is 
\begin{equation}
u=r_{0} \left[ \sqrt{ 1-\frac{2M}{r_{0}}-4\pi K\left( 2-\frac{\lambda }{r_{0}^{2}}\right)} \right]^{-1};  \label{impactu2}
\end{equation}
then, for photons coming from infinity such that their closest approach distance is the radius of the photon sphere (i.e. $r_{0}=r_{m}$), after some algebra we find that 
\begin{equation}
u_{m}=\frac{-16\pi K \lambda (1-8\pi K) +3M\left[ 3M+\sqrt{9M^{2}-32\pi K
\lambda (1-8\pi K)}\right] }{\sqrt{2(1-8\pi K)^{3}\left\{ -8\pi K \lambda
(1-8\pi K) +M\left[ 3M+\sqrt{9M^{2}-32\pi K \lambda (1-8\pi K)}\right]
\right\} }}.  \label{um2}
\end{equation}
Finally, from Eqs. (\ref{abarra1}) and (\ref{bbarra1}), the strong deflection limit coefficients for the black hole defined above are given by 
\begin{equation}
a_1=\sqrt{\frac{3M + \sqrt{9M^{2}-32\pi K \lambda (1-8\pi K)}}{2(1-8\pi
K)\sqrt{9M^{2}-32\pi K \lambda (1-8\pi K)}}}  \label{abarra2}
\end{equation}
and 
\begin{equation}
a_2=a_1\ln \left[ \frac{ -2\left( 1-8\pi K\right) r_{m}^{2}+16\pi
K\lambda }{2Mr_{m}-\left( 1-8\pi K\right) r_{m}^{2}-4\pi K\lambda }\right]
+I_{R}(r_{m})-\pi ,  \label{bbarra2}
\end{equation}
which are functions of the mass $M$, and the parameters of the model $K$ and $\lambda$. The deflection angle in the strong deflection limit is univocally determined for the Skyrmion black hole lens by replacing Eqs. (\ref{um2}), (\ref{abarra2}), and (\ref{bbarra2}) in Eq. (\ref{alpha}). Once the black hole and light source positions are determined, the angular positions of the relativistic images and their magnifications can be found by using Eqs. (\ref{pm14}), (\ref{pm15}), and (\ref{pm19}), while the observables defined in the previous section by Eqs. (\ref{ob1}), (\ref{ob4}), and (\ref{ob5}). In the case that $K=0$, the geometry (\ref{m2}) becomes Schwarzschild and the corresponding values $u_m^{\mathrm{Schw}} = 3 \sqrt{3} M$,  $a_1^{\mathrm{Schw}}=1$, and $a_2^{\mathrm{Schw}}=\ln [216(7-4\sqrt{3})]-\pi $ are recovered.

The expressions above are rather complicated. In order to understand their physical meaning, as we expect a small correction over the Schwarzschild geometry, it is useful to perform a first order Taylor expansion under the assumptions that $K \ll 1$ and $K\lambda/M^2 \ll 1$. In this case, the critical impact parameter is given by 
\begin{equation}
u_m = \left( 3 \sqrt{3} + 36 \sqrt{3} \pi K - \frac{2 \sqrt{3} \pi
K \lambda}{M^2} \right) M ,  \label{um3}
\end{equation}
while  the strong deflection limit coefficients take the form 
\begin{equation}
a_1= 1 + 4\pi K + \frac{4 \pi K \lambda}{9M^2}
\label{abarra3}
\end{equation}
and 
\begin{equation}
a_2= \ln [216(7-4\sqrt{3})]+ 4\pi K \ln [216(7-4\sqrt{3})] + \frac{4
\pi K \lambda}{9M^2} \left\{ \ln[216 (7 - 4 \sqrt{3})] + 2 \sqrt{3} - 6
\right\} -\pi .  \label{bbarra3}
\end{equation} 
The observables defined in the previous section adopt the form 
\begin{equation}
\theta_{\infty}= \left( 3 \sqrt{3} + 36 \sqrt{3} \pi K - 
\frac{2 \sqrt{3} \pi K \lambda}{M^2} \right) \frac{M}{D_{ol}},  \label{ob6}
\end{equation}
\begin{equation}
s= \left[ \frac{648 \sqrt{3} e^{-3 \pi}}{(2 + \sqrt{3})^2} + \frac{7776 \sqrt{3} e^{-3 \pi}(1 + \pi ) \pi K
} {(2 + \sqrt{3})^2 } + \frac{3888 e^{-3 \pi}(-7 -6 \sqrt{3} + 6 \pi + 4 
\sqrt{3} \pi ) \pi K \lambda} {9(2 + \sqrt{3})^3 M^2} \right] \frac{M}{D_{ol}
},  \label{ob7}
\end{equation}
and 
\begin{equation}
r= e^{2 \pi} - 8 \pi^2 e^{2 \pi} K - \frac{8 \pi ^2 e^{2 \pi} K
\lambda } {9M^2} .  \label{ob8}
\end{equation}
The first term in the right hand side of each equation is the value corresponding to the Schwarzschild geometry. 

In Planck units, using Eq. (\ref{window}), we have that $K=3.33 \times 10^{-41}$ and $0.0241\leq K\lambda \leq 0.0400$. The solar mass in these units is $M_{\odot } = 9.14 \times 10^{37}$; then, for a black hole with $M=10\ M_{\odot }$ we obtain $2.87 \times 10^{-80} \leq K \lambda /M^{2} \leq 4.78\times 10^{-80}$, while for the supermassive Galactic black hole with $M = 4 \times 10^{6}\ M_{\odot }$ we have that $1.80\times 10^{-91}\leq K\lambda /M^{2}\leq 2.99\times 10^{-91}$. So, in these cases, our first order Taylor expansion above is justified.  We see that for the Skyrmion black hole, the deviations of the strong deflection limit coefficients and observables from those corresponding to a Schwarzschild spacetime with the same mass in a possible astrophysical scenario are extremely small. 

On the other hand, being the present spherical black hole asymptotically locally flat, it is reasonable to compare it with another asymptotically locally flat black hole, the obvious candidate being the Barriola-Vilenkin black hole \cite{BVM}. This geometry can be recovered from Eq. (\ref{m2}) if we take $\lambda =0$ and we identify $K$ with the usual parameter $\eta ^2$. With these replacements, the equations above provide the strong deflection limit for the Barriola-Vilenkin spacetime, which was previously studied in Ref. \cite{cm}. To give a precise estimate of the mass of the Barriola-Vilenkin black hole is not easy (and, indeed, there is no common agreement in the literature on this issue). However, a natural order of magnitude for the mass of a black hole whose ``source" is a topological defect is around $10-100$ TeV (which is the order of magnitude for the gravitating topological defects appearing in the standard model, see, for instance Ref. \cite{choBH} and the references therein). In this case, the effects of the Skyrme term could become quite relevant compared with the Barriola-Vilenkin black hole. But the lensing distances for these small mass black holes should be very short, because the observables $\theta_{\infty}$ and $s$ are proportional to $M/D_{ol}$.

\section{Discussion}\label{discu}

We have extended the strong deflection limit to a class of spherically symmetric spacetimes which are asymptotically locally flat. From this logarithmic approximation for the deflection angle, we have presented the analytical expressions for the positions and the magnifications of the relativistic images, and for three useful observables. Although some asymptotically locally flat geometries were analyzed by using the strong deflection limit (e.g. \cite{cm}), a systematic approach was missing in the literature.

We have applied the formalism to a spherically symmetric black hole solution of the $SU(2)$ Einstein-Skyrme system. This model is of interest due to its close relation with the low energy limit of QCD. The metric possesses a solid angle deficit as the Barriola-Vilenkin black hole and a Reissner-Norsdr\"{o}m like term with a fixed positive constant replacing the square of the charge. To the best of authors knowledge, this is the first analytic derivation of the strong deflection limit in Einstein-Skyrme theory. We have analytically obtained the strong deflection limit coefficients, the positions and the magnifications of the relativistic images in terms of them, as well as the standard observables. 

We have also compared the strong deflection limit of the present spherical black hole in Einstein-Skyrme theory with those corresponding to the Schwarzschild and the Barriola-Vilenkin geometries. We have found that the deviations from the results corresponding to the Schwarzschild spacetime are extremely small (tens orders of magnitude) for a mass range from a few solar masses to supermassive objects, like the astrophysical black holes of interest. Consequently, the observation of these deviations is not expected with current or foreseeable future astronomical facilities. On the other hand, the deviations from the Barriola-Vilenkin black hole could be relevant when the black hole mass is of the typical order of magnitude of the masses of gravitating topological defects of the standard model \cite{choBH}. But gravitational lensing in this case will require very short lensing distances, i.e.  the presence of these very small size black holes under controlled conditions in a terrestrial laboratory. This is an intriguing possibility in view of the recent proposals on the production of black holes in particle accelerators \cite{eg}.

\section*{Acknowledgments}

This work has been supported by Universidad de Buenos Aires and by CONICET (E.F.E. and C.M.S.), and by the Fondecyt grant 1160137 (F.C.). The Centro de Estudios Cient\'{\i}ficos (CECS) is funded by the Chilean Government through the Centers of Excellence Base Financing Program of Conicyt (F.C.).

\end{document}